\documentclass[%
reprint,
superscriptaddress,
showpacs,preprintnumbers,
amsmath,amssymb,
 prb,
floatfix
]{revtex4-1}

\usepackage{bm}  
\usepackage{esint}
\usepackage{graphicx}
\usepackage{grffile} 
\usepackage{color}

\newcommand{\kk}{\mathbf{k}}
\newcommand{\qq}{\mathbf{q}}
\newcommand{\vecA}{\mathbf{A}}
\newcommand{\pp}{\mathbf{p}}

\begin{document}

\author{A.~F. Kemper}
\email{afkemper@lbl.gov}
\affiliation{Computational Research Division, Lawrence Berkeley National Laboratory, Berkeley, CA 94720, USA}
\affiliation{Stanford Institute for Materials and Energy Sciences, SLAC National Accelerator Laboratory, Menlo Park, CA 94025, USA}

\author{M. Sentef}
\affiliation{Stanford Institute for Materials and Energy Sciences, SLAC National Accelerator Laboratory, Menlo Park, CA 94025, USA}

\author{B. Moritz}
\affiliation{Department of Physics, Northern Illinois University, DeKalb, IL 60115, USA}
\affiliation{Stanford Institute for Materials and Energy Sciences, SLAC National Accelerator Laboratory, Menlo Park, CA 94025, USA}
\affiliation{Department of Physics and Astrophysics, University of North Dakota, Grand Forks, ND 58202, USA}

\author{C.~C. Kao}
\affiliation{Stanford Synchrotron Radiation Lightsource, SLAC National Accelerator Laboratory, Menlo Park, California 94025, USA}

\author{Z.~X. Shen}
\affiliation{Stanford Institute for Materials and Energy Sciences, SLAC National Accelerator Laboratory, Menlo Park, CA 94025, USA}
\affiliation{Geballe Laboratory for Advanced Materials, Stanford University, Stanford, California 94305, USA}

\author{J.~K. Freericks}
\affiliation{Department of Physics, Georgetown University, Washington, DC 20057, USA}

\author{T.~P. Devereaux}
\affiliation{Stanford Institute for Materials and Energy Sciences, SLAC National Accelerator Laboratory, Menlo Park, CA 94025, USA}
\affiliation{Geballe Laboratory for Advanced Materials, Stanford University, Stanford, California 94305, USA}

\title{Mapping of the unoccupied states and relevant bosonic modes via the time dependent momentum distribution}

\begin{abstract}
The unoccupied states of complex materials are difficult to measure, yet play a key role in determining their properties.
We propose a technique that can measure the unoccupied states, 
called time-resolved Compton scattering, which measures the time-dependent momentum distribution (TDMD).
Using a non-equilibrium Keldysh formalism, we study the TDMD for electrons coupled to a lattice in a pump-probe setup. 
We find a direct relation between temporal oscillations in the TDMD and the dispersion of
the underlying unoccupied states,
suggesting that both can be measured by time-resolved Compton scattering.
We demonstrate the experimental feasibility by applying the method to a model of MgB$_2$ with realistic
material parameters.
\end{abstract}
\maketitle

\section{introduction}
In understanding the emergent properties of complex materials, it is insufficient to
limit ones' study solely to the occupied electronic states.  The unoccupied states play an important
role in determining, for example, the nature of the gap in charge- and spin-density waves,
the absorption properties of semiconductors, or the magnetic properties
of Mott systems.  
For example, without knowledge of the unoccupied states, it is difficult to measure both
the size as well as the ordering wave vector $\vec q$ of a gap in the electronic spectrum.
Unfortunately, the most direct experimental measurements of
momentum-resolved 
states in quantum materials, angle-resolved photoemission spectroscopy (ARPES), 
only determines the occupied states.  This leaves the alternative methods of
inverse photoemission spectroscopy, which lacks the signal strength
to measure with sufficient resolution,
or indirectly
inferring the unoccupied dispersion from two-particle quantities such as optical
spectroscopy. 
Pump-probe spectroscopy, and in particular time-resolved (tr-) ARPES, can measure
part of the unoccupied states, but is limited by the inherent competition between temporal
and energy resolution.

Here we consider a different quantity, where one sacrifices the energy information in
favor of high time resolution, while maintaining high momentum resolution. 
We propose to measure the (gauge-invariant)
time-dependent momentum distribution (TDMD) of quasiparticles,
defined by $\langle c_\kk^\dagger(t) c_\kk(t)\rangle \equiv n_\kk(t)$, which is similar to
the so-called Wigner distribution.\cite{g_mahan}
It contains a rich amount of information
about the behavior of pumped quasiparticles while avoiding some of
the resolution-related complications of other measurements.

The TDMD is commonly measured via time-of-flight absorption images
in cold atomic gases, 
where it was recently used to map out the Fermi
surface of a gas of $^{40}$K atoms.\cite{t_drake_12}
In the context of condensed matter systems,
we propose two new types of measurements to access the TDMD. 
First, we propose to extend the technique of Compton scattering for solids into the time
domain\cite{[Time-resolved Compton scattering in 1D was previously studied by ]r_wagner_10}.  Compton scattering has a long history as a probe of the electron momentum
distribution for solids in equilibrium
\cite{m_cooper_99,a_bansil_01,y_sakurai_11}.
With the advent of x-ray free-electron laser sources, the photon energy of the ultrashort
x-ray probe pulses has been extended into the hard x-ray regime, where the fractional
Compton cross-section
becomes appreciable.
With the advent of a hard x-ray free electron laser at the Linac Coherent Light Source,
the photon energy will extend
 the energy of the x-ray free electron laser to beyond 20 keV.  
 There is also significant progress in crystal optics at these energies,
 such that time-resolved high resolution Compton scattering with momentum resolution
 of a few percent of the Brillouin zone will be feasible.
Additionally, recent developments in photonics suggest
the future availability of ultrashort pulses ranging from x-rays to gamma rays\cite{k_taphuoc_12}.
Time-resolved Compton scattering
has the advantage of using high-momentum photons, which can readily access
the Brillouin zone edges in all directions.
Secondly, one can access the TDMD with 
tr-ARPES by using similar time delays and probe pulse widths and 
integrating the tr-ARPES signal over energy (assuming the tr-ARPES signal
comes from a single band near the Fermi level).

\begin{figure*}[ht]
	\includegraphics[clip=true,trim=0 20 200 0,width=\textwidth]{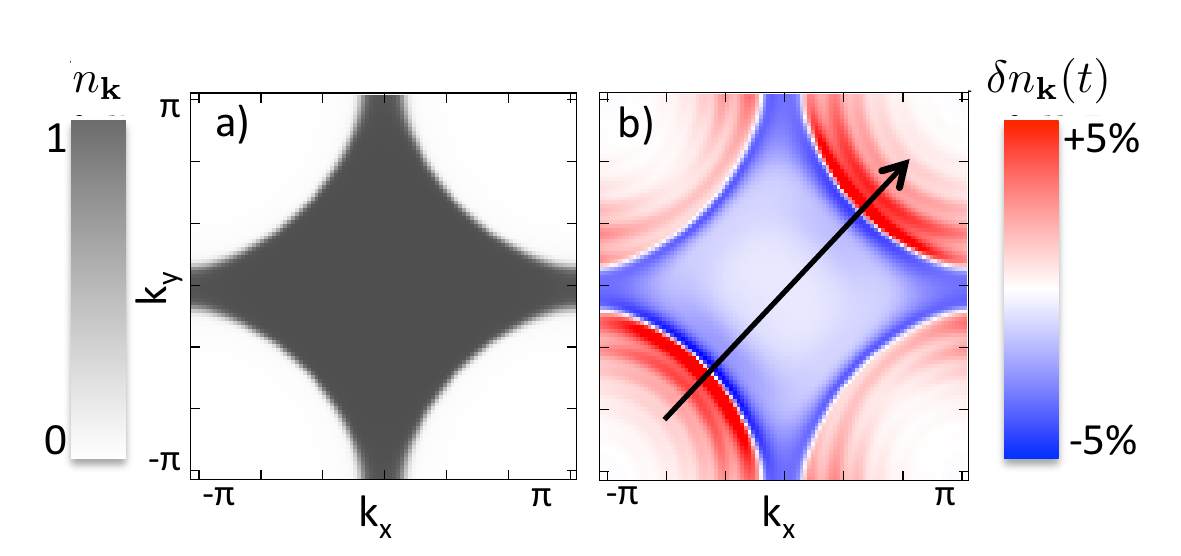}
	
	\caption{a) (Color online) Equilibrium TDMD $n_\kk(t)$ far before the pump arrives.  The gray
	area indicates where quasiparticle states are occupied.  The Fermi surface is sharply defined for the
	parameters presented ($g^2=\sqrt{0.02}V_{nn}, T=0.04V_{nn}, \Omega=0.4V_{nn}$).
	b) Change in the momentum distribution from equilibrium
	$\delta n_\kk(t)$ after a pump at $t\approx4.8\tau_p$.  The arrow indicates the direction of the applied
	electric field.}
	\label{fig:wigner2d}
\end{figure*}

We illustrate the use of the TDMD by 
demonstrating how it can be used to map out the unoccupied dispersion in a 
pump-probe setup on a model system with electron-lattice interactions
appropriate to a large class of complex materials.
The TDMD exhibits two ubiquitous phenomena
\textemdash
both due to electron-phonon interactions.
First, the phonons dissipate the energy delivered to
the electronic system by the pump.
Second, after pumping, oscillations related to the phonon frequencies
 (and independent of the initial pump parameters)
are commonly observed in the resulting spectra measured by the probe.
\cite{f_schmitt_08,l_perfetti_08,l_rettig_10,k_kim_12}
We will show that both of these features are clearly
seen in the TDMD, and that the oscillation frequencies can be
analyzed to extract the underlying dispersion, in particular for the
unoccupied states.
Finally, we will demonstrate the feasibility of this
technique by studying a model band structure for MgB$_2$.

The paper shall proceed as follows. In Sec.~\ref{sec:method}, we outline
our method for calculations of two-time Green's functions on the Keldysh contour.
In Sec.~\ref{sec:results}, we present our results for the TDMD of a model system,
and in Sec.~\ref{sec:mgb2}, we apply the technique to the case of MgB$_2$
to demonstrate realistic capabilities of our method to map unoccupied states
and their coupling to phonon spectra.  A summary is presented in Sec.~\ref{sec:summary}.

\section{Method}
\label{sec:method}
We study a two-dimensional system of electrons coupled to a bath
of non-dispersive phonons with frequency $\Omega$ via a constant coupling $g$
(known as the Holstein model),
\cite{t_holstein_59a,*t_holstein_59b}.
\begin{align}
\mathcal H = \sum_{\kk\sigma} (\epsilon(\kk) - \mu)  c^\dagger_{\kk,\sigma} c_{\kk,\sigma}+ \sum_\qq \Omega b^\dagger_\qq b_\qq \nonumber\\
 + g \sum_{\kk\qq\sigma} c^\dagger_{\kk+\qq,\sigma} c_{\kk,\sigma} \left( b_\qq + b^\dagger_{-\qq} \right)
\end{align}
where $c_\kk$ ($b_\qq$) annihilates an electron (phonon) of momentum $\kk$ ($\qq$).

While the TDMD is useful independent of the underlying model, here we
use a band structure motivated by 
the transition-metal oxides:
a 2D tight-binding $\epsilon(\kk)$ with (next) nearest-neighbor hoppings 
$V_{nn}$ and $V_{nnn} = 0.3 V_{nn}$,
and $\mu=-1.02V_{nn}$. We use the conventions $\hbar = e = c = 1$.
Below, we will report timescales in units of a characteristic phonon
timescale $\tau_p = 1/\Omega$.  For example, a phonon with an energy of
$\Omega=10$ meV has $\tau_p\approx 0.4$ ps.
Similarly, there is a characteristic
electron timescale $\tau_e = 1/V_{nn} \approx 16$ fs for $V_{nn}=250$ meV.
Although in this work we will explicitly examine higher energy phonons
for numerical stability reasons, the results below occur due to the relevant energy
scales in the problem, and will simply be rescaled to lower phonon energies.

To describe the non-equilibrium pump-probe process, we propagate the system on the
Kadanoff-Baym-Keldysh (Keldysh) contour\cite{kadanoff_baym,*l_keldysh_64},
which has been described in detail elsewhere
\cite{g_mahan,a_jauho_84,*j_davies_88,v_turkowski_05,*j_freericks_08}.
The non-equilibrium Keldysh formalism has been successfully used to describe the
time evolution of correlated electrons within dynamical mean-field theory
\cite{m_eckstein_08,*m_eckstein_09,*m_eckstein_09b,b_moritz_10,m_eckstein_10,b_moritz_11,*b_moritz_12}.
The electric field is included through the standard Peierls' substitution
$\kk \rightarrow \kk - \vecA(t)$, where the vector potential $\vecA$ is related to the applied
electric field $\mathbf{F}$ via $-\partial\vecA(t)/\partial t = \mathbf{F}$ and we work in the 
Hamiltonian gauge.  

Within the Migdal limit,
we consider the renormalization
of the electron Green's function by a single phonon emission/absorption and ignore the phonon
self-energy (as is done in conventional Migdal-Eliashberg theory). In this limit, we solve the Dyson
equation
\begin{align*}
G_\kk(t,t') = G^0_\kk(t,t') + \int_\mathcal{C} dt_1 dt_2 G^0_\kk(t,t_1) \Sigma(t_1,t_2) G_\kk(t_2,t')
\end{align*}
with the self-energy $\Sigma(t,t') = i g^2 \sum_\kk D^0(t,t') G^0_{\kk}(t,t')$,
where $G^0_\kk(t,t')$ and $D^0(t,t')$ are the 
non-interacting
electron and phonon Green's functions, respectively
\cite{[For a complete description of the non-interacting Green's functions see e.g. ]g_mahan}.
The Dyson equation can be solved by recasting it as a matrix equation\cite{j_freericks_08}
or by decomposing into Volterra-type equations through the Langreth rules\cite{r_van_leeuwen_05},
which is the approach used here.

\section{Results}
\label{sec:results}
Figure~\ref{fig:wigner2d} shows the gauge-invariant TDMD 
$n_\kk(t)=iG_{\kk+\vec A(t)}^<(t,t'\rightarrow t)$
for $g = \sqrt{0.02}V_{nn}$, $\Omega = 0.4V_{nn}$ and at an initial temperature $T=0.04 V_{nn}$.
Fig.~\ref{fig:wigner2d} (a) shows $n_\kk(t)$ at times far before the pump, where the system is fully described
by the equilibrium problem.
To clearly show the changes for the small pump fluences considered 
($F_{max} \le 4V_{nn}$),
Fig.~\ref{fig:wigner2d} (b) shows the change in the TDMD from equilibrium:
$\delta n_\kk(t) \equiv n_\kk(t) - n_\kk(t\rightarrow -\infty)$.

We apply a pulse centered at time $t=0$ in the $(11)$ (diagonal) direction of the form 
$A(t) = \left(F_\mathrm{max}/\omega_A\right) \exp(-t^2/(2\sigma^2)) \sin(\omega_A t)$,
with a maximum field strength of $F_\mathrm{max}=4V_{nn}$, a frequency
$\omega_A=2V_{nn}$ and $\sigma \approx 0.4 \tau_p$.  
At the time shown in Fig.~\ref{fig:wigner2d} (b), the pump has passed, and the system is relaxing
towards equilibrium.  The $\mathcal C_4$ symmetry of $n_\kk(t)$ in equilibrium is broken by
the field (which points in the $(11)$ direction as shown by the black arrow),
and this is reflected in the transient change.  In fact, this can be used to emphasize particular
regions of interest in the Brillouin zone.  By rotating the field, the regions exhibiting the largest
change in $n_\kk(t)$ will shift with the field direction, providing access to new regions
by emphasizing the experimental signal near different momenta.

\begin{figure}[h]
	\includegraphics[clip=true,trim=35 20 125 80,width=\columnwidth]{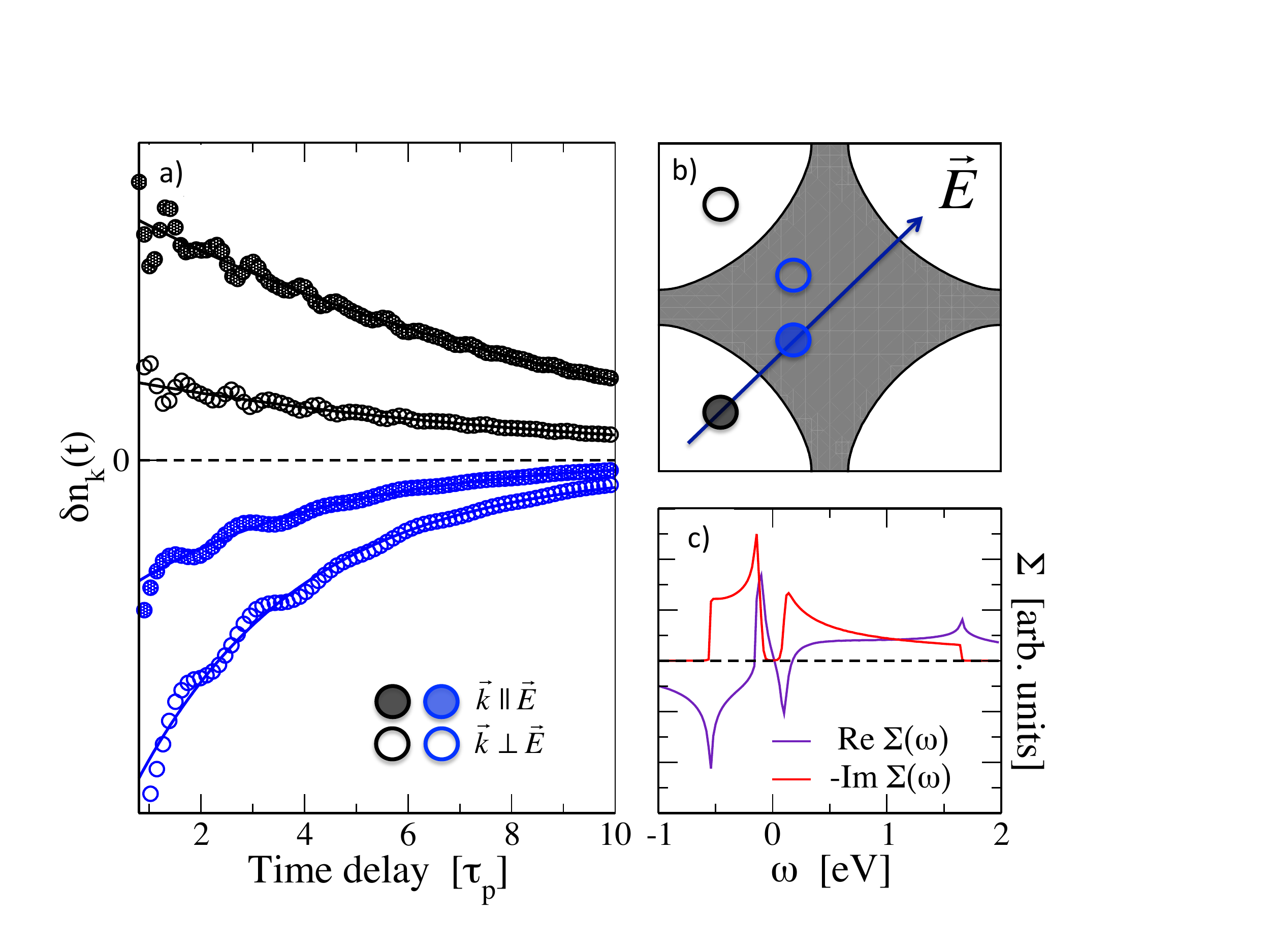}
	\caption{a) (Color online) Change in the TDMD $n_\kk(t)$ from the equilibrium value (at $t=0$) for
	the points in the Brillouin zone indicated in b).  Symbols are calculated data points (for clarity, 
	not all are shown), solid
	lines are fits to an exponential decay
	using a fixed relaxation rate $1/\tau_\kk = -2\textrm{Im}\ \Sigma(\omega=\epsilon_\kk)$.
	c)  Real and imaginary parts of the equilibrium self-energy $\Sigma(\omega)$ 
	for the parameters in a) ($g=\sqrt{0.02}V_{nn}$, $T=0.04V_{nn}$, $\Omega=0.4V_{nn}$).}
	\label{fig:decayfits}
\end{figure}

First, we consider the anisotropic redistribution of quasiparticles as a function of time.
In Figure~\ref{fig:decayfits} (a), we show $\delta n_\kk(t)$ at the momenta
indicated in the inset;
the blue (black) momenta
have the same energy $\epsilon_\kk$ but due to the $\mathcal C_4$ symmetry breaking 
have different time traces (Fig.~\ref{fig:decayfits} (b)).  
From the figure, one can clearly see that there is
a difference in the overall magnitude of the change at the equivalent momenta (with the same color).
The solid lines are fits to the amplitude of a simple decaying
exponential, where we have taken the decay constant at a momentum $\kk$ 
from the on-shell \textit{equilibrium} self-energy
$1/\tau_\kk(\omega)=-2\textrm{Im}\ \Sigma(\omega=\epsilon_\kk)$ shown in 
Fig.~\ref{fig:decayfits} (c),
as suggested by a recent study\cite{m_sentef_12}.
The remarkable
agreement with the calculated data points confirms that it is indeed the equilibrium self-energy that
controls the decay rate. 
Since the equilibrium self-energy is $\mathcal C_4$ symmetric, the anisotropic redistribution
in $\delta n_\kk(t)$ is caused by the $\mathcal C_4$ symmetry breaking of the applied pump field.

\begin{figure}[h]
	\includegraphics[clip=true,trim=37 50 350 40,width=\columnwidth]{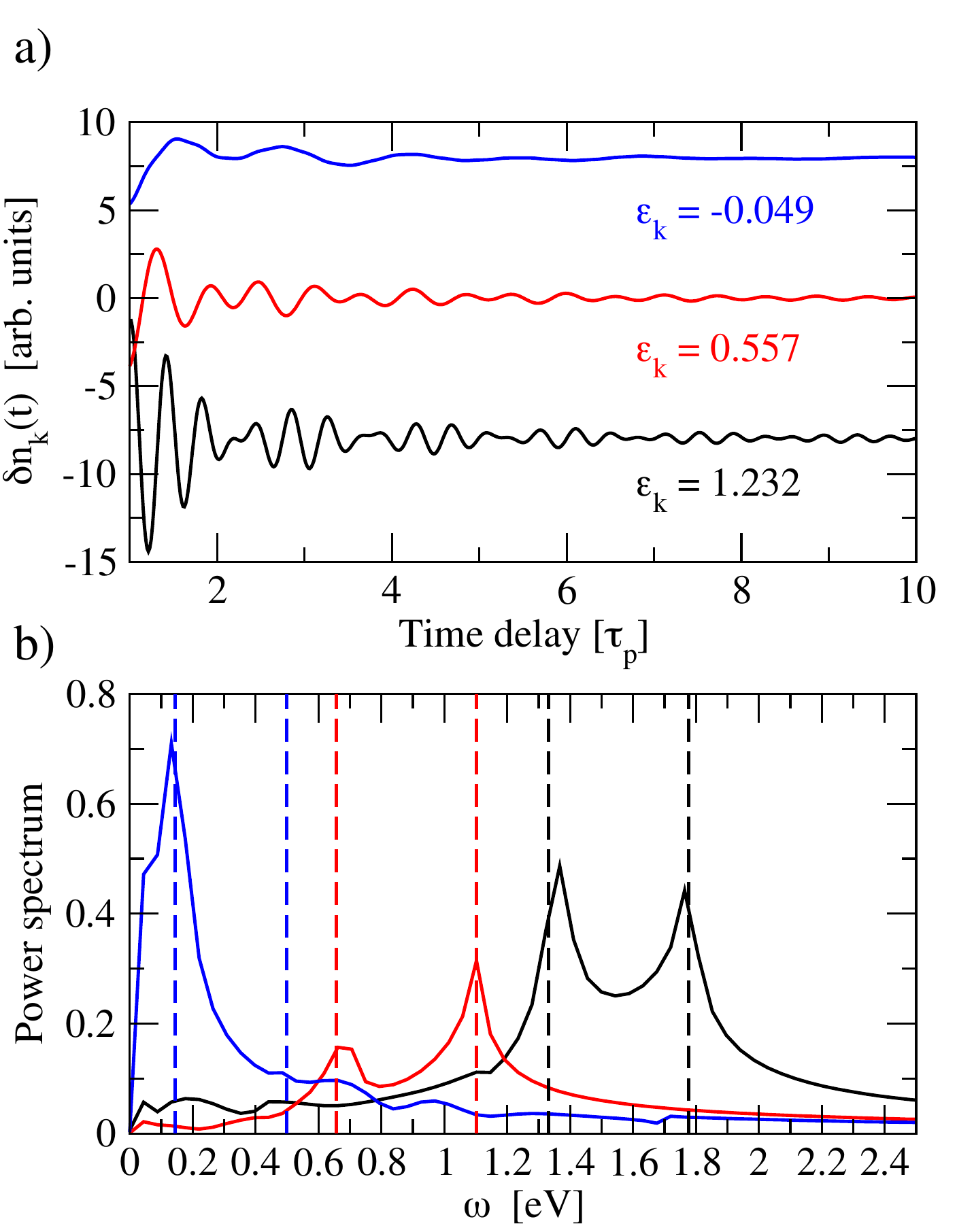}
	\caption{a) (Color online) Oscillatory part of $\delta n_\kk(t)$ for $\kk$ along (11) at the energies
	$\epsilon_\kk$ indicated (offset for clarity).
	b) Fourier transform of a).  Vertical lines indicate the frequencies corresponding
	to the strong oscillations on the unoccupied side,
	$\omega = \epsilon_\kk + \Omega$ and $\omega = \epsilon_\kk -W_- + \Omega$.
	In both a) and b) the oscillations on the unoccupied side close to the Fermi level (blue
	curves) have been scaled down by 4 for visibility. 
	(see text for details).}
	\label{fig:oscillations}
\end{figure}

On top of the decay are oscillations dominated by a few characteristic frequencies,
as often observed in pump-probe experiments.
\cite{f_schmitt_08,l_perfetti_08,l_rettig_10,k_kim_12}
These are clearly observed as rings in
$\delta n_\kk(t)$ shown in Fig.~\ref{fig:wigner2d}, as well as oscillations in
the time traces in Fig.~\ref{fig:decayfits} (a).   We further emphasize these in
Fig.~\ref{fig:oscillations}, where we have subtracted the decaying exponential 
from several points along the Brillouin zone diagonal in Fig.~\ref{fig:wigner2d}.
From Fig.~\ref{fig:oscillations} (a), one can observe two main features.  First, the
oscillation frequencies in the time traces depend strongly on momentum.
Second, although there is only a single phonon frequency in our model,
several oscillation frequencies can be observed.
Figure ~\ref{fig:oscillations} (b) shows the Fourier transform power spectrum of the oscillations in 
Fig.~\ref{fig:oscillations} (a).
Each curve has readily visible maxima, in addition to some smaller structure across the frequency
spectrum.  For each energy (momentum) $\epsilon_\kk$, the vertical lines indicate the energies
$\omega = \epsilon_\kk + \Omega$ and $\omega=\epsilon_\kk - W_- + \Omega$, where $W_-$ is the
energy at the bottom of the band (at $\kk = \Gamma$).

\begin{figure}[htpb]
	\includegraphics[clip=true,trim=25 50 80 90,width=0.9\columnwidth]{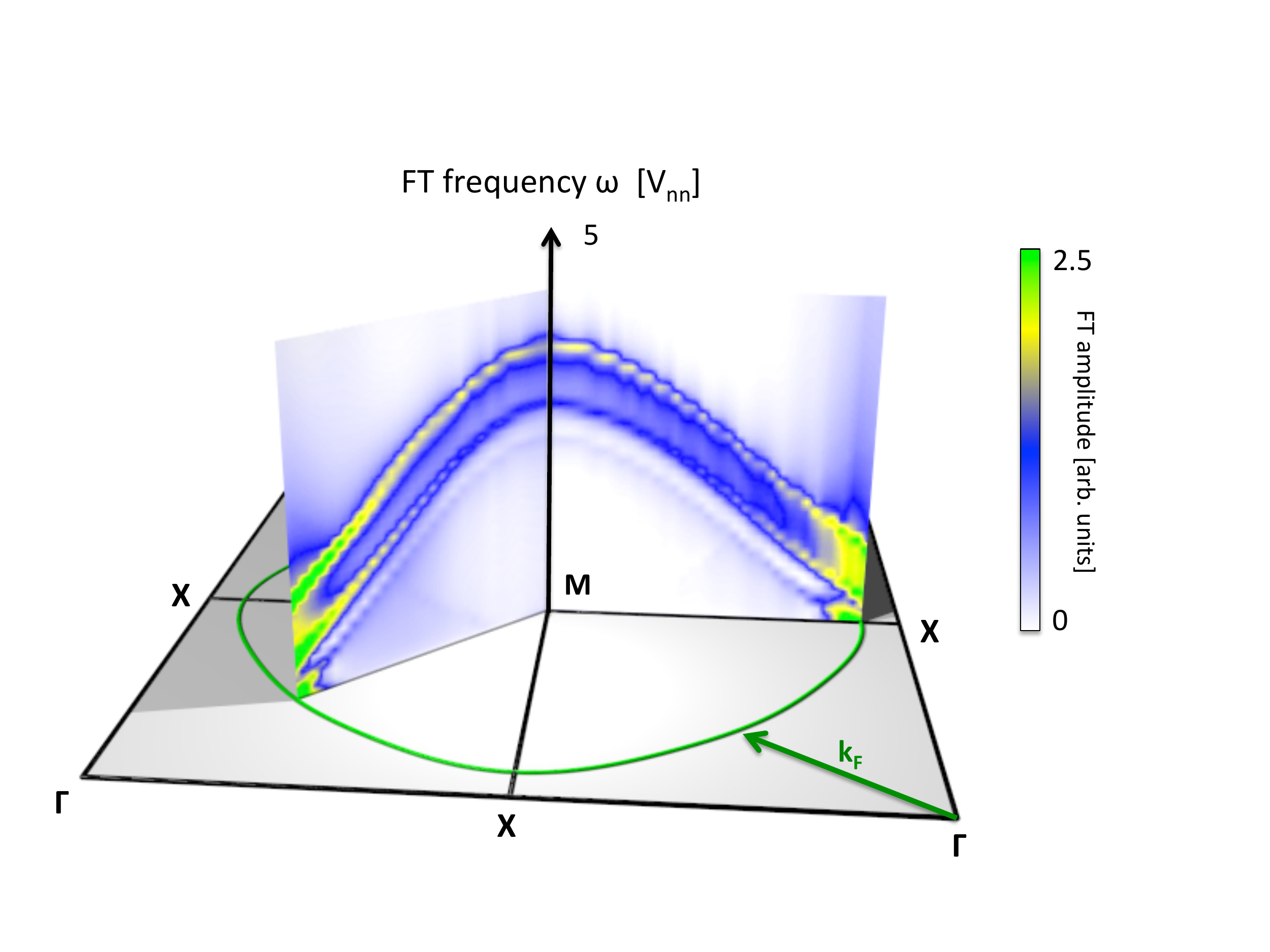}
	
	\caption{(Color online) Fourier power spectrum of the oscillation frequencies along the
	$\Gamma$-M and M-X directions
	in the unoccupied Brillouin zone.  The two strong peaks in the Fourier spectra follow the lines
	corresponding to phonon emission processes; a small tertiary line of peaks can be seen at lower frequencies
	corresponding to a phonon absorption process. The Fermi surface is shown in green on the base
	plane.}
	\label{fig:ft_powerspec}
\end{figure}

The temporal dynamics of the Holstein model leads to this rich physics.
The decay and the oscillations observed are momentum (energy) dependent, and their amplitudes
depend on the electron-phonon coupling, as well as the
applied field.  
Describing the full temporal dynamics of $n_\kk(t)$ is complicated, yet we can gain some
insight by focusing on the times where the pump is off, while the system
is still displaced from its equilibrium configuration.  The subsequent relaxation occurs
from the new configuration through the equations of motion. 
In equilibrium, the TDMD is balanced by equal rates
of in- and out-scattering;
once the pump breaks this symmetry, the system
relaxes back to the equilibrium according to the dynamics contained
within the self-energy and Green's functions.  In addition to the relaxation,
the TDMD oscillates at the band energy shifted by the characteristic
energies in the problem (the phonon frequency and bandwidth) at which
the self-energy is largest. 
This is shown analytically at $T=0$ in the appendix; the full calculation requires the numerical methods used here.

At low temperatures,  
the real part of the self-energy $\Sigma^\prime$ is large and peaked near four frequencies,
$\omega = \pm \Omega$ and $\omega = W_\pm \pm \Omega$,
where $W_\pm$ is the upper (lower) band edge
(see Fig.~\ref{fig:oscillations}(b)).
The resulting $n_\kk(t)$ oscillates at the band energy $\epsilon_\kk$
shifted by those frequencies.
In principle, all four frequencies should be observable in the power spectrum of
Fig.~\ref{fig:oscillations}; however, two frequencies (with the positive sign above) correspond
to phonon absorption, which is small at low temperatures, and thus the corresponding 
peaks in the power spectrum are reduced significantly.
The two strong peaks expected for the unoccupied side
are indicated in Fig.~\ref{fig:oscillations} by vertical lines, which agree well with the observed frequencies,
up to a small shift in energy due to $\Sigma^\prime(\omega=0)$ which is normally
absorbed into the chemical potential $\mu$.  
An interesting consequence of the
expected oscillation frequencies is that if one considers a measurement near the Fermi level,
the oscillations will appear to be strongest at just the phonon frequency (as in
recent tr-ARPES experiments\cite{f_schmitt_08,l_perfetti_08,l_rettig_10,k_kim_12}).

The simple dependence of the oscillation frequencies on the underlying band structure suggests
a novel method to measure the electron dispersion by looking at the
oscillations in the TDMD.
The oscillations are most clearly
visible in the unoccupied region of the Brillouin zone, which is exactly the region that
traditional methods
for measuring the dispersion have difficulty accessing.
By tracking the oscillation frequencies as a function of momentum
one can directly map out the dispersion.
In Fig.~\ref{fig:ft_powerspec},
we plot the Fourier transform power spectrum in the unoccupied portion of the Brillouin
zone along the zone diagonal and zone face.  This shows most clearly the two strong peaks
in the Fourier transform associated with phonon emission, although a weaker line corresponding
to phonon absorption is also visible below the two main lines.  All three lines,
up to the constant shifts 
from $\Sigma^\prime(0)$,$W_\pm$ and $\Omega$, follow the unoccupied dispersion.
Thus, from the maxima in the power spectrum the unoccupied dispersion can be directly
measured.

The high temporal resolution afforded by the neglect of any frequency information
(in contrast with tr-ARPES) allows these oscillations to be clearly resolved.
Furthermore, the oscillations are strongest in the direction of the applied field.
This gives the method an additional degree of freedom, where the field direction
can be used to select the momentum cuts of interest.  Measuring the TDMD thus provides
complementary information to that gained from other time-resolved experiments.
Additionally, we have shown that there is a direct connection between the oscillations
in the TDMD and $\Sigma^\prime(\omega)$.  This is the complement to the recent work
by Sentef et al.\cite{m_sentef_12}, where it was shown that the
\textit{imaginary} part of the self-energy $\Sigma^{\prime\prime}(\omega)$ can be
inferred from
decay rates in tr-ARPES.

\section{A case study: M\lowercase{g}B$_2$}
\label{sec:mgb2}
\begin{figure*}[ht]
	\includegraphics[width=0.38\textwidth]{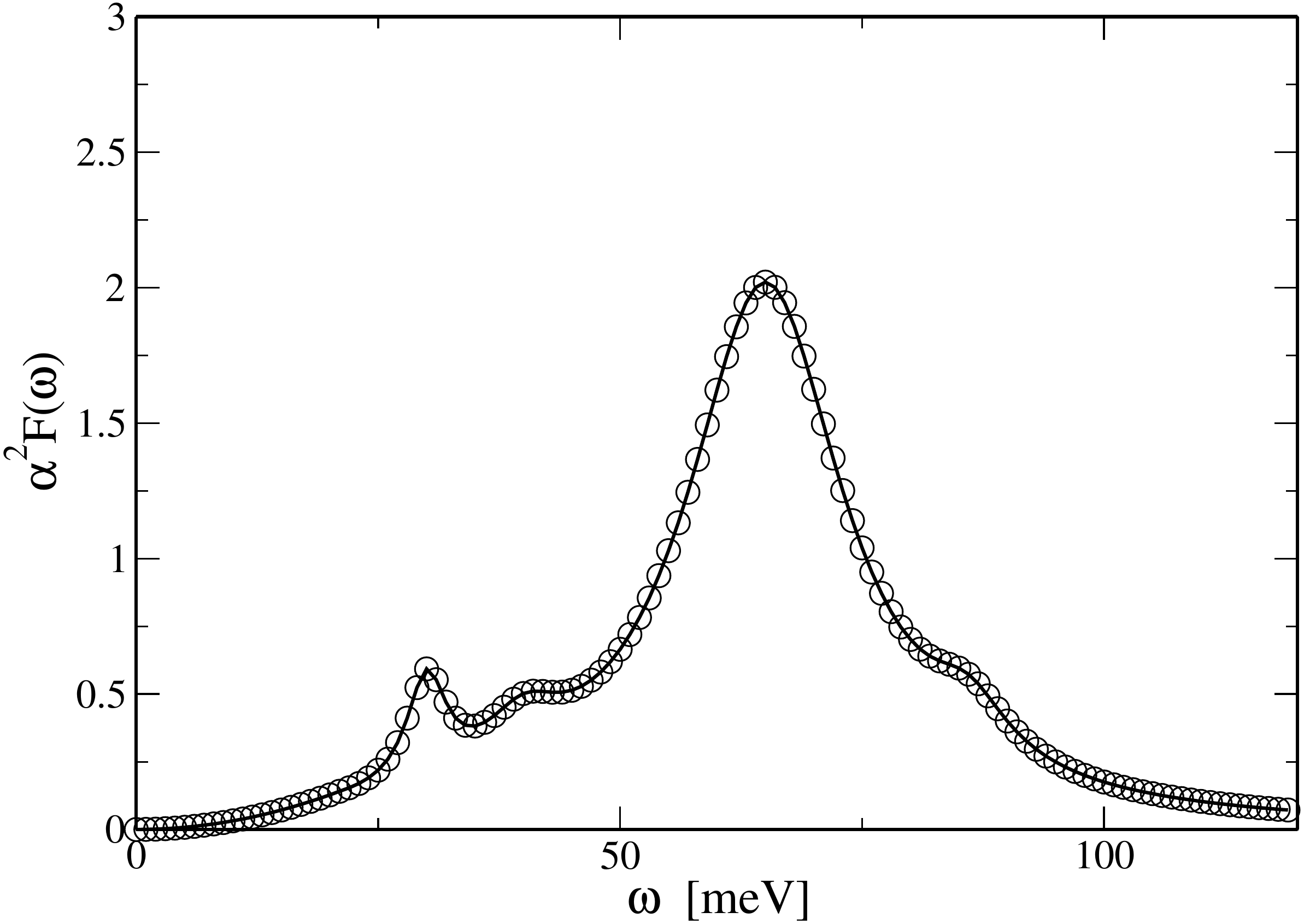}
	\includegraphics[width=0.61\textwidth]{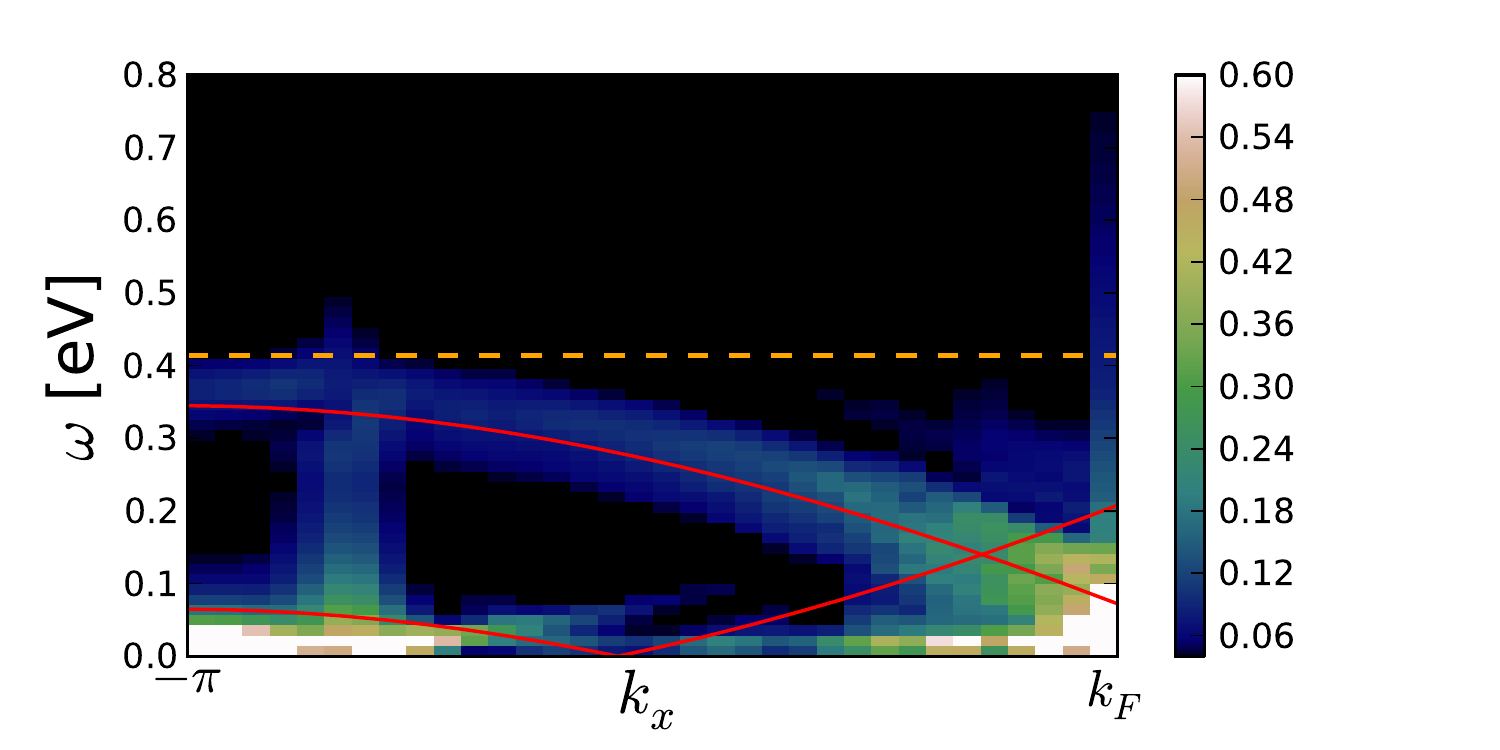}
	
	\caption{(Color online) 
	Left: Model $\alpha^2F(\omega)$ for MgB$_2$ used in the calculations.
	Right: Fourier power spectrum of the oscillation frequencies near the $\pi$ bonding band in MgB$_2$.
	The red lines indicate the expected frequencies due to emission of a phonon of 65 meV, and the orange
	line indicates an resolution upper bound for a temporal resolution of 10 fs.
	}
	\label{fig:ft_powerspec_mgb2}
\end{figure*}
The above analysis, applied to theoretical results, can be applied to experimental
results equally well. Here, the experimental capabilities must be considered. The salient
points to consider are as follows. First, the experimental time resolution will place an
upper bound on the oscillation frequencies that can be resolved.  This will limit the measurement capability
to regions near the Fermi level; fortunately, this is generally where one wishes to measure, and
thus the experimental resolution only limits the upper bound of resolvability.  Second, materials
have more than a single Holstein phonon mode, which should be taken into account.  In this section,
we repeat the calculations done previously, but for a band structure appropriate to MgB$_2$.\cite{k_szalowski_06}
Furthermore,
we couple the electronic system to a distribution of phonons through a
model $\alpha^2F(\omega)$, which we have constructed based on
experimentally broadened phonon frequencies calculated
from first-principles.\cite{k_bohnen_01}

Fig.~\ref{fig:ft_powerspec_mgb2} shows the model $\alpha^2F(\omega)$ used in the calculations.  For calculations
of the TDMD, we focus on the
$\pi$ bonding band of MgB$_2$, along the line $k_y=0, k_z=0$ of the band structure
\begin{align}
 \epsilon_\pi(\kk) =& e_\pi + 2t_\perp \cos k_z \nonumber \\
 &- t^\prime_{||}\sqrt{1+4\cos \frac{k_y}{2}\left(\cos \frac{k_y}{2}+\cos\frac{k_x\sqrt{3}}{2}\right)},
\end{align}
where $e_\pi=0.04$ eV, $t_\perp=0.92$ eV, and $t^\prime_{||}=1.60$ eV, and the Fermi level is set to 0.\cite{k_szalowski_06}

In principle, the redistribution of quasiparticles during the pump in a multi-band
system is complex; however, here we focus entirely on the decay of the excitations after the pump, which are determined
by the equilibrium, intra-band self-energy. Furthermore, our intent here is to include realistic material properties to
show that this signal can, in fact, be measured.  
As such, we perform the calculations for the $\pi$ bonding band only.
Since the self-energy is a priori unknown (unlike in the model calculation),
we have fitted an exponential to the time traces.
The fitted exponential
is subtracted from the time traces, 
which are subsequently Fourier transformed.  
The Fourier transform power spectrum
is shown as a false-color intensity map for a cut along the zone boundary in Fig.~\ref{fig:ft_powerspec_mgb2}. The red
lines on the plot indicate the expected oscillation frequencies for a single mode at $\Omega=65$ meV, 
namely (the larger) $\omega_1 = |\epsilon(\kk) + \Omega|$ and (near the Fermi level)
$\omega_2 = |\epsilon(\kk) + \Omega+W|$, where $W$ is the top
of the band. Although a full phonon distribution is included through $\alpha^2F(\omega)$, the presence of a strongly
coupled mode allows for the resolution of a single dispersive mode, although broadened.  The agreement with the
predicted frequencies and the observed ones indicates that our previous conclusions hold, even in the presence of
multiple phonon modes. The observation that the power spectrum is dominated by the single
broadened mode for a realistic phonon spectrum justifies the use of a single Holstein
mode previously, as it captures the essential features seen here.
The final, orange dashed line on the plot indicates the experimental resolution bound
for a probe resolution of 10 fs.

The further experimental considerations involve scattering matrix elements, and the distinction between
real and crystal momentum.  These issues have been discussed in some detail (in equilibrium) previously,
and the arguments presented there hold in non-equilibrium as well.\cite{w_schulke_96,m_cooper_99}.\\

\section{Summary}
\label{sec:summary}
We have shown the use of the time-dependent momentum distribution as a novel concept in the understanding
of time-resolved spectroscopy, and non-equilibrium phenomena in general.  The decay rates and the oscillations
in the time traces of the TDMD were shown to be directly related to the underlying equilibrium self-energy.  This
result is of importance in time-resolved spectroscopy, where it has long been assumed that the equilibrium properties
can be studied in a time-resolved experiment.  We further have shown that the TDMD can be directly measured by an experiment
that only has momentum and time resolution, such as time-resolved Compton scattering and energy-integrated time-resolved
ARPES.  
Experimentally, the width of the probe pulse limits the temporal resolution of the TDMD.
However, in many systems of interest, 
the features one wishes to investigate lie near the Fermi level, both above and below.  
For example, in the high-Tc cuprates, several features are predicted to lie within this energy range.\cite{b_moritz_09}  
In other gapped systems, without the knowledge of the spectra above the Fermi level
it is impossible to correctly assign the gap magnitude.  Similar effects due to spin fluctuations, phonons, and other interesting collective and bosonic features lie at low energy, and are thus
accessible by this technique.

\acknowledgments{\textit{Acknowledgments} 
AFK, MS, BM and TPD were supported by the U.S. Department of Energy, 
Basic Energy Sciences, Materials Sciences and Engineering Division under 
contract No. DE-AC02-76SF00515. 
JKF was supported by the U.S. Department of Energy, 
Basic Energy Sciences, Materials Sciences and Engineering Division under contract No. 
DE-FG02-08ER46542 and by the McDevitt bequest at Georgetown University. 
The collaboration was supported by the U.S. Department of Energy, Basic Energy Sciences, 
Materials Sciences and Engineering Division under contract Nos. DE-FG02-08ER46540
and DE-SC0007091.
This work was made possible by the resources of the National Energy Research Scientific
Computing Center 
which is supported by the U.S. DOE, Office of Science, under Contract No. DE-AC02-05CH11231.
We gratefully acknowledge discussions with P.~S. Kirchmann, J. Sobota, M. Wolf, and S. Yang.}

\appendix*

\section{Analytic calculation of oscillations}

We can explain the oscillations in the TDMD by examining the equations of motion for
the mixed and lesser Green's functions at $T=0$ (where the state is fully described without
a thermal average).
Following the Langreth rules, we obtain the following Volterra-type equations\cite{r_van_leeuwen_05},
\begin{widetext}
\begin{align}
\left[i\partial_t - h(t)\right]G_\pp^{ri}(t,0) =& \int_0^t d\bar t\ \Sigma^R(t,\bar t) G_\pp^{ri}(\bar t,\tau)
	\label{eq:Gri}\\
G_\pp^{ri}(0,0) =&\  i n_\pp \nonumber \\
\left[i\partial_t - h(t)\right] G_\pp^<(t,t') =& \int_0^t d\bar t\ \Sigma^R(t,\bar t) G_\pp^<(\bar t,t') 
   +\int_0^{t'} d\bar t\ \Sigma^<(t,\bar t) G_\pp^A(\bar t,t') 
      \label{eq:glesser}\\
   G_\pp^<(0,t') =& -G_\pp^{ri}(t',0)^* \nonumber
\end{align}
\end{widetext}
where $h(t) = \epsilon(\kk-\vecA(t))$, $\epsilon(\kk)$ is the dispersion, and $n_\pp$ is the Fermi function
at energy $\epsilon_\pp$.  The superscripts $<$ and $ri$
denote the ``lesser'' and mixed components
of the Green's function and self-energy, respectively.	

In the absence of a driving field or interactions, we obtain the known results for the mixed and lesser
Green's functions
\begin{align*}
G_{0\pp}^{ri}(t,0) =& i n_\pp e^{-i \epsilon_\pp t}\\
G_{0\pp}^<(t,t') =& i n_\pp e^{-i\epsilon_\pp(t-t')}
\end{align*}
We are interested in the oscillations of the TDMD after the pump is off, when the system
is out of its equilibrium state.  Since we are interested in the dynamics, in lieu of driving
the system explicitly with a field, we turn on the interactions at an infinitesimal time after
zero ($t=0^+$). By construction, this problem has the same dynamics as those of the
relaxation after driving by a pump. 
This, coupled with the approximation of weak interactions (where we replace
the full Green's function on the right hand side with the bare one), allows
us to obtain analytic formulae which show the oscillation dynamics.
We shall consider the change of the equilibrium Green's functions $\Delta G = G - G_0$.
First, we solve equation of motion for $\Delta G^{ri}(t)$, which we shall need as an initial
condition for the TDMD.  $\Delta G^{ri}(t)$ has no initial condition beyond that for the bare
Green's function $G_0^{ri}(t)$.
\begin{align}
\Delta G_\pp^{ri}(t,0) =& i g^2 n_\pp \sum_\kk \bigg[ 
\frac{n_\kk}{(\epsilon_\kk - \Omega - \epsilon_\pp)^2} F(-\Omega) \nonumber \\
&+ \frac{1-n_\kk}{(\epsilon_\kk+ \Omega - \epsilon_\pp)^2} F(\Omega) \bigg] \\
 \nonumber \\
F(\Omega) = e^{-i(\epsilon_\kk + \Omega)t}  &+ e^{-i \epsilon_\pp t} \big[i(\epsilon_\kk +\Omega - \epsilon_\pp)t-1 \big] \nonumber
\end{align}
We proceed by similarly solving the equation for the lesser component.  The full expressions
are complex, but can be significantly simplified by considering the case where $t' \rightarrow t$,
\begin{widetext}
\begin{align}
\Delta G^<_\pp(t,t') =& -G^{ri}(t)^* e^{-i\epsilon_\pp t} + i g^2 n_\pp \sum_\kk \bigg[
\frac{n_\kk}{(\epsilon_\kk - \Omega - \epsilon_\pp)^2} F(-\Omega) 
+ \frac{1-n_\kk}{(\epsilon_\kk + \Omega - \epsilon_\pp)^2} F(\Omega)
- \frac{ 2\cos\left[ (\epsilon_\kk - \Omega - \epsilon_\pp) t\right] - 2}{(\epsilon_\kk - \Omega + \epsilon_\pp)^2}
\bigg] \nonumber \\
=& 2 i g^2 n_\pp \sum_\kk (1-n_\kk) \bigg[ 
\frac{ \cos\left[ (\epsilon_\kk + \Omega - \epsilon_\pp)t\right]-1}{(\epsilon_\kk + \Omega - \epsilon_\pp)^2}
-\frac{ \cos\left[ (\epsilon_\kk - \Omega - \epsilon_\pp)t\right]-1}{(\epsilon_\kk - \Omega - \epsilon_\pp)^2}
\bigg]
\end{align}
\end{widetext}

Now, only the integral over the $\kk$ states remains. We shall assume that
the density of states is flat over the region
we integrate over (and equal to $N_0$), and has band edges at $\pm W$. 
This results in
integrals of the form
\begin{align}
\int_0^W& dx\ \frac{ \cos\left[(x\pm\Omega - \epsilon_\pp)t\right]-1}{(x\pm\Omega-\epsilon_\pp)^2}=\nonumber\\
&\frac{\cos\left[(x\pm\Omega - \epsilon_\pp)t\right]-1}{x\pm\Omega - \epsilon_\pp} 
- t \mathcal{S}\left[(x\pm\Omega - \epsilon_\pp)t\right] \bigg|_0^W
\end{align}
Here, $\mathcal{S}$ denotes the sine integral.  Thus, the two terms each show two distinct
oscillation frequencies,
$\omega = \epsilon_\pp \pm \Omega$ and $\omega = \epsilon_\pp \pm (\Omega +W)$.
\begin{figure}
	\includegraphics[width=\columnwidth]{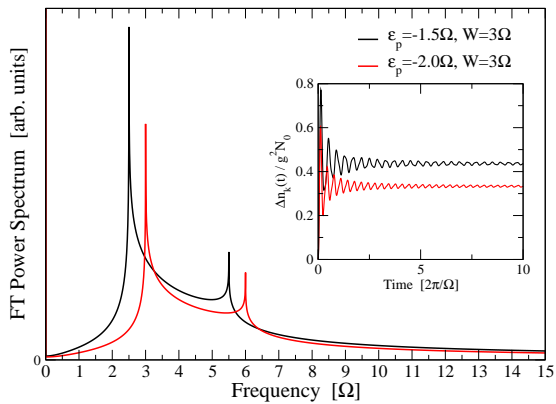}
	\caption{Change in the TDMD after turning on interactions for an infinite band.}
	\label{fig:deltank}
\end{figure}
Figure~\ref{fig:deltank} shows the change in the TDMD as the interactions are turned on in an
infinite band.  Here, the oscillations occur at $\epsilon_\pp \pm \Omega$, although those
where $|\epsilon_\pp - \Omega|$ is smallest dominate the signal.

In the course of this calculation, a number of approximations were made.  Nevertheless, the oscillation
frequencies agree with those observed from the full numerical simulations.  The inclusion of
the correct density of states, as well as the replacement of the bare Green's functions will
cause a small shift of the frequencies observed, and a distribution of frequencies
centered around the bare one shown here.

\bibliography{wigner,timedomain}

\end{document}